\documentclass[11pt]{article}
\usepackage[utf8]{inputenc}
\usepackage[T1]{fontenc}
\usepackage{amsmath,amssymb,bm}
\usepackage{geometry}
\usepackage{cite}
\newcommand{\ellNL}{\ell_{\rm NL}}
\newcommand{\LambdaG}{\Lambda_G}
\newcommand{\Rin}{\mathcal{R}_{\rm in}}
\newcommand{\Rph}{\mathcal{R}_{\rm ph}}
\newcommand{\Tph}{\mathcal{T}_{\rm ph}}

\title{Conditions for Gravitational-Wave Echo Suppression\\
in Diffeomorphism-Invariant Nonlocal Quantum Gravity}
\author{J. W. Moffat\\
Perimeter Institute for Theoretical Physics, Waterloo, Ontario N2L 2Y5, Canada\\
and\\
Department of Physics and Astronomy, University of Waterloo, Waterloo,\\
Ontario N2L 3G1, Canada}
\date{July 27, 2026}

\begin{document}
\maketitle

\begin{abstract}
Gravitational-wave echoes can arise when perturbations are trapped between distinct scattering structures in the effective potential of an ultracompact merger remnant. We examine the conditions under which analytic nonlocal smearing can reduce such signals in a Gaussian-smeared regular compact-object geometry. For configurations possessing a true event horizon, the standard purely ingoing boundary condition excludes an inner echo cavity independently of the ultraviolet completion. For horizonless configurations, smoothness alone does not eliminate echoes. The axial Regge--Wheeler potential can contain a trapping region bounded by an inner centrifugal barrier and an outer potential maximum. We distinguish echoes caused by the global geometric potential from reflection by an additional localized inner transition layer. If such a layer is smeared over a proper radial length $\ellNL$ and the metric function is approximately constant across it, its tortoise-coordinate width is enhanced by the gravitational redshift. In the weak-scattering approximation, a Gaussian layer produces the conditional suppression. This result requires a localized weak interaction, slow variation of the background across the layer, and a valid Born approximation. It is not a model-independent exclusion of echoes from smooth horizonless objects. The analysis identifies the redshift, nonlocal-length, and potential-structure conditions that must be evaluated before gravitational-wave observations can be converted into constraints on analytic nonlocal quantum gravity.
\end{abstract}

\section{Introduction}
\label{sec:introduction}

It was first hypothesized in 2016 that gravitational wave echoes could exist if there was a reflective boundary for black hole horizons~\cite{GW1,GW2,GW3,GW4}. Searches for statistically significant post-merger echoes have reported no evidence for them~\cite{Abac,Westerweck,Liu2026}. Model-dependent and model-agnostic searches likewise constrain the amplitude or signal-to-noise ratio of possible late-time components rather than excluding all exotic compact-object models. Consequently, observational non-detections constrain a model only after its microscopic parameters have been mapped to an effective potential, a frequency-dependent inner response, and a waveform.

Analytic nonlocal theories use entire functions of the covariant d'Alembertian to soften ultraviolet behavior, while retaining diffeomorphism invariance~\cite{Moffat1990,EMKW1991,Moffat2011,Moffat2016,ModestoMoffatNicolini2011,Biswas2012,GreenMoffat2021,MoffatThompson2026}. In this paper the term form factor denotes the analytic function appearing in the action or field equations. The same object is called a regulator when emphasizing its ultraviolet damping role. The two terms refer to the same analytic construction but to different aspects of it.

We use units $c=\hbar=1$ and metric signature $(-,+,+,+)$. Gravitational-wave observations of compact-binary mergers provide direct access to strong-field gravity. The post-merger ringdown is commonly described by the quasinormal modes of the remnant, and departures from the Kerr spectrum can test modified gravity and near-horizon structure.

A gravitational-wave echo is a delayed component of the response produced when part of a perturbation is temporarily trapped and subsequently leaks to infinity. The phase-dependent echo criterion for this same Gaussian-smeared family, including the distinction between an outer-horizon endpoint at \(r_*=-\infty\) and a regular horizonless center at finite \(r_*\), was
analyzed in. A frequently studied mechanism is repeated scattering between the photon-sphere barrier and an inner reflecting surface or transition layer~\cite{Cardoso2016,CardosoPani2017,AbediDykaarAfshordi2017}. More generally, echoes can be generated by a smooth multi-peak effective potential containing a trapping well; a discontinuous surface or nonanalytic boundary condition is not necessary. 

We consider an ultracompact horizonless object to be strongly redshifted, when its lapse has a positive but small minimum, $f_{\min}\ll1$. By an inner transition region, we mean a finite proper-distance interval in which the effective source, equation of state, or geometry changes from an exterior black-hole-like regime to a regular interior regime. This transition need not be a material surface and need not be reflective.

\section{Analytic nonlocal framework and Gaussian-smeared geometry}
\label{sec:background}

A representative curvature-quadratic nonlocal action is:
\begin{equation}
S=\frac{1}{16\pi G}\int d^4x\sqrt{-g}\left[
 R+R\mathcal{F}_1(\Box)R+R_{\mu\nu}\mathcal{F}_2(\Box)R^{\mu\nu}
 +C_{\mu\nu\rho\sigma}\mathcal{F}_3(\Box)C^{\mu\nu\rho\sigma}
 \right],
\label{eq:nonlocal_action}
\end{equation}
where the $\mathcal{F}_i$ are analytic functions chosen, so that the propagator is softened in the Euclidean ultraviolet without introducing additional zeros in the entire-function dressing factor around the background of interest~\cite{Biswas2012,GreenMoffat2021,KoshelevMazumdar2017,BuoninfanteMazumdar2019}. The diffeomorphism invariant operator $\Box$ is defined by $\Box=g^{\mu\nu}\nabla_\mu \nabla_\nu$.

A typical Euclidean damping factor is:
\begin{equation}
 \mathcal{E}(p_E^2)=\exp\!\left[-\left(\frac{p_E^2}{\LambdaG^2}\right)^n\right],
 \qquad n\geq1,
\label{eq:entire_factor}
\end{equation}
with $\ellNL\equiv\LambdaG^{-1}$. The interpretation of an entire-function nonlocality length as an irreducible response or localization width, rather than as a lattice spacing or spacetime discretization, was developed in~\cite{ThompsonLocalization2026,ThompsonLocalizationb2026}.

Expanding an entire form factor at $|\Box|\ll\LambdaG^2$ gives an infinite derivative expansion. Keeping only finitely many terms can reproduce local curvature-squared corrections, but such a truncation is not dynamically equivalent to the untruncated theory, because it changes the pole structure~\cite{MoffatThompson2026,KouroshniaNuclearClock2026}. It can share a low-energy curvature expansion with local quadratic gravity, while differing in its propagating degrees of freedom. Horizonless ultracompact objects in local quadratic gravity provide an important comparison class~\cite{SalvioVeermae2020}, but their perturbation and echo properties cannot be transferred to the nonlocal theory without a separate calculation.

The present perturbation analysis is linear in the metric perturbation. The form factor is not expanded in powers of $\Box/\LambdaG^2$ when deriving the smearing kernel. A separate Born approximation is introduced only for a localized weak inner interaction in Sec.~\ref{sec:localized_layer}. We do not analyze nonlinear merger dynamics or a nonperturbative strong-coupling regime. 

To display a concrete regular geometry, we use the effective-source construction of Refs.~\cite{Moffat2011,Moffat2016,ModestoMoffatNicolini2011}. 
A distributional source is replaced by:
\begin{equation}
 T_{\mu\nu}(x)\longrightarrow T^{\rm eff}_{\mu\nu}(x)
 =\mathcal{F}\!\left(\frac{\Box}{\LambdaG^2}\right)T_{\mu\nu}(x),
\label{eq:effective_source}
\end{equation}
where $\mathcal{F}(0)=1$. In a weakly curved static approximation, the choice $\mathcal{F}=\exp(\Box/\LambdaG^2)$ acts as a spatial heat kernel. We emphasize that the metric below is a solution of Einstein equations sourced by this effective stress tensor. It is used as a controlled background model and is not asserted here to be an exact solution of the full action~\eqref{eq:nonlocal_action}.

Consider:
\begin{equation}
 ds^2=-f(r)dt^2+\frac{dr^2}{f(r)}+r^2d\Omega^2,
 \qquad
 T^{\mu}{}_{\nu}=\mathrm{diag}[-\rho,p_r,p_t,p_t].
\label{eq:sss_metric}
\end{equation}
Writing:
\begin{equation}
 f(r)=1-\frac{2Gm(r)}{r},\qquad m'(r)=4\pi r^2\rho(r),
\label{eq:mass_definition}
\end{equation}
we take the normalized Gaussian density:
\begin{equation}
 \rho(r)=\frac{M}{(4\pi\ellNL^2)^{3/2}}
 \exp\!\left(-\frac{r^2}{4\ellNL^2}\right).
\label{eq:gaussian_density}
\end{equation}
The mass function and lapse are then:
\begin{align}
 m(r)&=M\left[\operatorname{erf}\!\left(\frac{r}{2\ellNL}\right)
 -\frac{r}{\sqrt{\pi}\ellNL}\exp\!\left(-\frac{r^2}{4\ellNL^2}\right)\right],
\label{eq:mass_function}\\
 f(r)&=1-\frac{2GM}{r}\left[\operatorname{erf}\!\left(\frac{r}{2\ellNL}\right)
 -\frac{r}{\sqrt{\pi}\ellNL}\exp\!\left(-\frac{r^2}{4\ellNL^2}\right)\right].
\label{eq:lapse_function}
\end{align}
At fixed $GM$, varying $\mu=GM/\ellNL$ is equivalent to varying the nonlocal length. Equation~\eqref{eq:mass_function} shows directly that every profile approaches $m(r)/M\to1$ as $r/\ellNL\to\infty$, while the small-$r$ expansion below gives the regular $m(r)\propto r^3$ core.

For $r\ll\ellNL$:
\begin{equation}
 m(r)=\frac{4\pi}{3}\rho_0r^3+\mathcal{O}(r^5),\qquad
 f(r)=1-\frac{8\pi G\rho_0}{3}r^2+\mathcal{O}(r^4),
\label{eq:de_sitter_core}
\end{equation}
where $\rho_0=M/(4\pi\ellNL^2)^{3/2}$. This particular Gaussian model has a regular de Sitter expansion near the origin. This is a property of the model, not a general theorem that every regularization of a Schwarzschild singularity must produce a de Sitter core.

Introduce:
\begin{equation}
 x=\frac{r}{\ellNL},\qquad \mu=\frac{GM}{\ellNL},\qquad
 A(x)=\operatorname{erf}\!\left(\frac{x}{2}\right)-\frac{x}{\sqrt{\pi}}e^{-x^2/4}.
\label{eq:dimensionless_variables}
\end{equation}
Then, we have
\begin{equation}
 f(x)=1-\frac{2\mu}{x}A(x).
\label{eq:dimensionless_lapse}
\end{equation}
The horizon equation depends only on $\mu$, because all lengths have been scaled by $\ellNL$. The extremal configuration satisfies $f(x_\star)=f'(x_\star)=0$, giving numerically:
\begin{equation}
 x_\star=3.02244,\qquad \mu_\star=1.90412.
\label{eq:critical_values}
\end{equation}
For $\mu>\mu_\star$ there are two positive roots of $f(x)=0$. For $\mu=\mu_\star$ there is one degenerate positive root, and for $\mu<\mu_\star$ there is no positive root. The complete horizon phase structure follows directly from Eqs.~\eqref{eq:dimensionless_lapse} and~\eqref{eq:critical_values}.

Conservation of the effective stress tensor gives:
\begin{equation}
 p_r'+\frac{f'}{2f}(\rho+p_r)+\frac{2}{r}(p_r-p_t)=0.
\label{eq:stress_conservation}
\end{equation}
The commonly used choice $p_r=-\rho$ determines $p_t$ through Eq.~\eqref{eq:stress_conservation}. The perturbation analysis below must nevertheless state an additional assumption about perturbations of this effective matter sector.

\section{Axial perturbations and the complete local potential}
\label{sec:axial_potential}

We first analyze local wave propagation on the fixed background~\eqref{eq:sss_metric}. We restrict attention to axial gravitational perturbations with spherical harmonic index $L\geq2$ and assume that the axial perturbation of the effective matter source vanishes. This assumption closes the gravitational master equation. A full derivation from the nonlocal action may modify both the kinetic operator and the source response.

With time dependence $e^{-i\omega t}$, the master variable obeys:
\begin{equation}
 \left[\frac{d^2}{dr_*^2}+\omega^2-V_L^{\rm axial}(r)\right]\Psi_{\omega L}=0,
 \qquad \frac{dr_*}{dr}=\frac{1}{f(r)},
\label{eq:rw_equation}
\end{equation}
where, for the metric~\eqref{eq:sss_metric}:
\begin{equation}
 V_L^{\rm axial}(r)=f(r)\left[
 \frac{L(L+1)}{r^2}-\frac{f'(r)}{r}+\frac{2[f(r)-1]}{r^2}
 \right].
\label{eq:axial_potential}
\end{equation}
For $f=1-2GM/r$, Eq.~\eqref{eq:axial_potential} reduces to the Schwarzschild Regge--Wheeler potential $f[L(L+1)/r^2-6GM/r^3]$~\cite{ReggeWheeler1957,Chandrasekhar1983}.

In dimensionless variables:
\begin{equation}
 \ellNL^2V_L^{\rm axial}(x)=f(x)\left[
 \frac{L(L+1)}{x^2}-\frac{1}{x}\frac{df}{dx}
 +\frac{2[f(x)-1]}{x^2}\right].
\label{eq:dimensionless_potential}
\end{equation}
For the two-horizon and extremal cases, the exterior $L=2$ potential obtained from Eq.~\eqref{eq:dimensionless_potential} has the usual outer barrier and the horizon imposes an ingoing boundary condition. In the near-critical horizonless case $\mu=1.80$, however, direct numerical evaluation of the same equation shows that the regular-center centrifugal rise and an outer maximum produce a potential well. The well has a local minimum near:
\begin{equation}
 x_{\min}\simeq3.216,\qquad \ellNL^2V_{\min}\simeq0.0228,
\end{equation}
and an outer maximum near:
\begin{equation}
 x_{\max}\simeq5.841,\qquad \ellNL^2V_{\max}\simeq0.0468.
\end{equation}
Modes in the appropriate frequency band can therefore be temporarily trapped and leak outward as delayed components.

A true-horizon solution has no inner outgoing component when the standard condition:
\begin{equation}
 \Psi_{\omega L}\sim e^{-i\omega r_*},\qquad r_*\rightarrow-\infty,
\label{eq:horizon_condition}
\end{equation}
is imposed. This conclusion does not rely on ultraviolet smearing. A horizonless solution instead has a regular center. Regularity implies $\Psi_{\omega L}\propto r^{L+1}$ near $r=0$, but it does not imply zero reflection from the complete interior potential. The relevant question is whether Eq.~\eqref{eq:axial_potential} contains a trapping structure, and this must be checked case by case. 

\section{Nonlocal smearing of a localized inner transition layer}
\label{sec:localized_layer}

We now consider a different and more limited question: how does analytic smearing alter reflection from an additional localized inner layer? Let $\rho$ be proper radial distance:
\begin{equation}
 d\rho=\frac{dr}{\sqrt{g(r)}},
\label{eq:proper_distance}
\end{equation}
and consider the general static metric $ds^2=-fdt^2+dr^2/g+r^2d\Omega^2$. Since:
\begin{equation}
 dr_*=\frac{dr}{\sqrt{f(r)g(r)}}=\frac{d\rho}{\sqrt{f(r)}},
\label{eq:tortoise_proper_relation}
\end{equation}
a layer of fixed proper thickness has a redshift-enhanced width in tortoise coordinate.

Suppose a localized interaction is centered at $r=r_0$, and that $f(r)\simeq f_0>0$ across the layer. A normalized Gaussian of proper width $\ellNL$ corresponds to the tortoise-coordinate width:
\begin{equation}
 \sigma_*=\frac{\ellNL}{\sqrt{f_0}}.
\label{eq:tortoise_width}
\end{equation}
We model its contribution to the master potential by:
\begin{equation}
 W_{\rm NL}(r_*)=\frac{\lambda}{\sqrt{\pi}\sigma_*}
 \exp\!\left[-\frac{(r_*-r_{*0})^2}{\sigma_*^2}\right],
 \qquad \int_{-\infty}^{\infty}W_{\rm NL}(r_*)dr_*=\lambda.
\label{eq:gaussian_layer}
\end{equation}
For this specific kernel:
\begin{equation}
 \widetilde{W}_{\rm NL}(k)=\lambda e^{ikr_{*0}}e^{-k^2\sigma_*^2/4}.
\label{eq:gaussian_transform}
\end{equation}
The behavior of the Fourier transform is derived explicitly rather than postulated as a frequency filter.

For an isolated weak interaction, the first Born reflection amplitude is:
\begin{equation}
 \mathcal{R}_{\rm B}(\omega)=\frac{1}{2i\omega}
 \int_{-\infty}^{\infty}dr_*\,e^{2i\omega r_*}W_{\rm NL}(r_*).
\label{eq:born_amplitude}
\end{equation}
Using Eq.~\eqref{eq:gaussian_transform}:
\begin{equation}
 \mathcal{R}_{\rm B}^{\rm NL}(\omega)=
 \frac{\lambda}{2i\omega}e^{2i\omega r_{*0}}
 \exp[-\omega^2\sigma_*^2]
 =\mathcal{R}_{\rm B}^{(0)}(\omega)
 \exp\!\left[-\frac{\omega^2\ellNL^2}{f_0}\right].
\label{eq:conditional_suppression}
\end{equation}
A static observer at $r_0$ measures:
\begin{equation}
 \omega_{\rm loc}(r_0)=\frac{\omega}{\sqrt{f_0}},
\label{eq:local_frequency}
\end{equation}
so Eq.~\eqref{eq:conditional_suppression} can be written:
\begin{equation}
 \mathcal{R}_{\rm B}^{\rm NL}(\omega)=\mathcal{R}_{\rm B}^{(0)}(\omega)
 \exp[-\omega_{\rm loc}^2(r_0)\ellNL^2].
\label{eq:local_suppression}
\end{equation}
The appearance of $\omega_{\rm loc}$ is a consequence of defining the smearing width in proper distance and then converting to the tortoise coordinate. It is not obtained by replacing $\omega$ with $\omega_{\rm loc}$ in a global flat-space form factor.

The Born approximation requires a separate smallness condition. A sufficient one-dimensional criterion is:
\begin{equation}
 \eta_{\rm B}=\frac{1}{2\omega}\int_{-\infty}^{\infty}|W_{\rm NL}(r_*)|dr_*
 =\frac{|\lambda|}{2\omega}\ll1,
\label{eq:born_condition}
\end{equation}
together with the self-consistency requirement $|\mathcal{R}_{\rm B}|\ll1$. Equation~\eqref{eq:conditional_suppression} cannot be used to prove small reflection from an initially strong or nearly perfectly reflecting wall, while simultaneously relying on the Born approximation. In that regime the full scattering problem must be solved.

Substantial suppression requires:
\begin{equation}
 \omega_{\rm loc}\ellNL=\frac{\omega\ellNL}{\sqrt{f_0}}\gtrsim1,
 \qquad\hbox{or}\qquad f_0\lesssim(\omega\ellNL)^2.
\label{eq:suppression_condition}
\end{equation}
This condition displays the dependence on the nonlocal scale. For a stellar-mass remnant and a Planck-length $\ellNL$, one has parametrically $\omega\ellNL\sim10^{-40}$, so appreciable suppression would require $f_0$ of order $10^{-80}$ or smaller. A larger nonlocal length weakens this redshift requirement. The assertion that suppression is independent of the detailed nonlocal scale is not retained.

At a true horizon, static observers do not extend through the horizon, and the divergence of Eq.~\eqref{eq:local_frequency} should not be used as an independent proof of vanishing reflection. The correct exterior statement is the ingoing boundary condition~\eqref{eq:horizon_condition}. Equation~\eqref{eq:local_suppression} applies instead to a horizonless layer with $f_0>0$, or to a localized layer outside a horizon, under the assumptions stated above.

\section{Echo transfer function and time-domain smearing}
\label{sec:transfer}

Let $\Rph(\omega)$ and $\Tph(\omega)$ denote the reflection and transmission amplitudes of the outer photon-sphere barrier, and let $\Rin(\omega)$ denote the response of a specified inner structure. A standard cavity transfer function is:
\begin{equation}
 \widetilde{Z}_{\rm echo}(\omega)\propto
 \frac{\Tph(\omega)\Rin(\omega)e^{2i\omega L}}
 {1-\Rph(\omega)\Rin(\omega)e^{2i\omega L}},
\label{eq:transfer_function}
\end{equation}
where $L$ is the one-way cavity length in tortoise coordinate. We use $\Rph$ consistently for the outer barrier; there is no separate $\mathcal{R}_{\rm ps}$ notation.

For a Planckian nonlocal energy scale, the asymptotic ringdown frequency satisfies $\omega_{\rm QNM}\ll\LambdaG$. At the photon sphere there is no extreme redshift enhancement, so a Planck-width smearing factor is essentially unity. We therefore do not multiply the photon-sphere reflectivity by an appreciable ultraviolet filter. If the only nonlocal modification is the localized inner layer of Sec.~\ref{sec:localized_layer}, then:
\begin{equation}
 \Rin^{\rm NL}(\omega)\simeq\Rin^{(0)}(\omega)e^{-\omega^2\sigma_*^2}
\label{eq:inner_response}
\end{equation}
within the Born regime, while $\Rph$ is unchanged at leading order. The $n$th return is consequently reduced by:
\begin{equation}
 |\Rph\Rin^{\rm NL}|^n=|\Rph\Rin^{(0)}|^n
 e^{-n\omega^2\sigma_*^2}.
\label{eq:nth_echo}
\end{equation}
This is a conditional bound for the chosen layer model, not a universal Paley--Wiener bound on every echo-producing geometry.

Multiplication by $e^{-\omega^2\sigma_*^2}$ corresponds in the time domain to convolution with:
\begin{equation}
 K_t(t)=\frac{1}{2\sqrt{\pi}\sigma_*}
 \exp\!\left(-\frac{t^2}{4\sigma_*^2}\right).
\label{eq:time_kernel}
\end{equation}
An already existing pulse is broadened on the tortoise-time scale $\sigma_*$ and its peak is reduced. This statement follows directly from the Gaussian convolution kernel and is not presented as a numerical waveform prediction for the Gaussian-smeared compact object.

The full horizonless problem cannot in general be reduced to Eq.~\eqref{eq:inner_response}. If the geometry itself produces the potential well identified by Eqs.~\eqref{eq:dimensionless_potential} and the numerical extrema quoted in Sec.~\ref{sec:axial_potential}, the wave equation must be evolved with the complete potential, including any nonlocal modification of the perturbation operator and effective stress response. Smearing one localized contribution does not remove other turning points elsewhere in the integration domain.

\section{Observational interpretation and limitations}
\label{sec:observations}

Current LVK non-detections constrain phenomenological echo amplitudes, delays, damping factors, or long-lived mode amplitudes. They do not directly give a bound on $\LambdaG$. Such a bound requires the chain:
\begin{equation}
 \LambdaG\longrightarrow
 \{f(r),g(r),\delta T^{\rm eff}_{\mu\nu},\mathcal{O}_{\rm pert}\}
 \longrightarrow V_L^{\rm NL}(r,r')
 \longrightarrow \Rin(\omega)
 \longrightarrow h_{\rm echo}(t),
\label{eq:parameter_mapping}
\end{equation}
where $\mathcal{O}_{\rm pert}$ denotes the full linearized nonlocal operator. The present paper supplies a background diagnostic and an analytic layer estimate, but not this complete mapping. We do not quote an event-by-event numerical constraint on the form factor.

Independent laboratory probes of an entire-function nonlocality scale have also been proposed using the \(^{229}\mathrm{Th}\) nuclear-clock transition and nonlocal time--energy response widths~\cite{KouroshniaNuclearClock2026}. Such matter-sector bounds cannot be identified directly with \(\Lambda_G\) unless the theory specifies a relation between the matter and gravitational nonlocality scales.

The principal conclusions are that for the true-horizon branch, the standard ingoing boundary condition removes the inner echo cavity unless the theory creates an additional reflector outside the horizon. For the horizonless Gaussian branch, smoothness does not guarantee the absence of echoes. Near criticality the axial potential can contain a trapping well. A separately specified weak inner layer is exponentially less reflective, when its redshift-enhanced tortoise width satisfies $\omega\ellNL/\sqrt{f_0}\gtrsim1$. The amount of suppression depends explicitly on $\ellNL$, the redshift $f_0$, the integrated layer strength, and the validity of the weak-scattering approximation.

A complete time-domain calculation from the linearized nonlocal equations is needed to determine which of these mechanisms dominates a physical merger remnant.

\section{Conclusions}
\label{sec:conclusions}

We have investigated echo suppression in analytic nonlocal quantum gravity. Analyticity and smoothness alone do not prevent a horizonless compact object from producing echoes. For the Gaussian-smeared geometry studied here, the near-critical horizonless branch develops a trapping region in the axial Regge--Wheeler potential. This provides an explicit example, within the paper's own background model, of how a smooth geometry can support delayed leakage.

For a remnant with a true horizon, the absence of an inner echo cavity follows from the purely ingoing horizon boundary condition and does not require a local-frequency Born argument. For a horizonless remnant, nonlocal smearing can suppress reflection from an additional localized transition layer, but only conditionally. A layer of proper thickness $\ellNL$ located where the lapse is $f_0$ has tortoise width $\ellNL/\sqrt{f_0}$; in the Born approximation its reflection amplitude acquires the factor $\exp[-\omega^2\ellNL^2/f_0]$. This derivation explains the role of the locally measured frequency, while making the required assumptions explicit.

The absence of detected echoes is compatible with analytic nonlocal quantum gravity, but it does not by itself establish that nonlocality has erased all possible trapping structures. Conversely, a future echo detection would constrain the complete nonlocal remnant model, not simply falsify every analytic entire-function form factor. Quantitative tests require the full linearized nonlocal perturbation equations and time-domain waveforms that map the nonlocal scale and remnant parameters into observable echo amplitudes and delays.

\section*{Acknowledgments}
I thank Ethan Thompson for helpful discussions. Research at the Perimeter Institute for Theoretical Physics is supported by the Government of Canada through Industry Canada and by the Province of Ontario through the Ministry of Research and Innovation.

\end{document}